# Temperature dependence of the electrical resistivity and the anisotropic magnetoresistance (AMR) of electrodeposited Ni-Co alloys

B.G. Tóth[a,+,*], L. Péter[a], Á. Révész[b], J. Pádár[a] and I. Bakonyi[a]

[a]*Research Institute for Solid State Physics and Optics, Hungarian Academy of Sciences.*
*H-1525 Budapest, P.O.B. 49, Hungary*

[b]*Department of Materials Physics, Eötvös University.*
*H-1518 Budapest, P.O.B. 32, Hungary*

**Abstract** ─ The electrical resistivity and the anisotropic magnetoresistance (AMR) was investigated for Ni-Co alloys at and below room temperature. The Ni-Co alloy layers having a thickness of about 2 μm were prepared by electrodeposition on Si wafers with evaporated Cr and Cu underlayers. The alloy composition was varied in the whole concentration range by varying the ratio of Ni-sulfate and Co-sulfate in the electrolyte. The Ni-Co alloy deposits were investigated first in the as-deposited state on the substrates and then, by mechanically stripping them from the substrates, as self-supporting layers both without and after annealing. According to an X-ray diffraction study, a strongly textured face-centered cubic (fcc) structure was formed in the as-deposited state with an average grain size of about 10 nm. Upon annealing, the crystal structure was retained whereas the grain size increased by a factor of 3 to 5, depending on alloy composition. The zero-field resistivity decreased strongly by

---

[+]Ph.D. student at Eötvös University, Budapest, Hungary
[*]Corresponding author. E-mail: tothb@szfki.hu

annealing due to the increased grain size. The annealing hardly changed the AMR below 50 at.% Co but strongly decreased it above this concentration. The composition dependence of the resistivity and the AMR of the annealed Ni-Co alloy deposits was in good quantitative agreement with the available literature data both at 13 K and at room temperature. Both transport parameters were found to exhibit a pronounced maximum in the composition range between 20 and 30 at.% Co and the data of the Ni-Co alloys fit well to the limiting values of the pure component metals (fcc-Ni and fcc-Co). The only theoretical calculation reported formerly on fcc Ni-Co alloys yielded at $T = 0$ K a resistivity value smaller by a factor of 5 and an AMR value larger by a factor of about 2 than the corresponding low-temperature experimental data, although the theoretical results properly reproduced the composition dependence of both quantities.

## 1. Introduction

More than twenty years after the discovery of the phenomenon of giant magnetoresistance (GMR) in nanoscale ferromagnetic/non-magnetic (FM/NM) metallic multilayers [1,2], the anisotropic magnetoresistance (AMR) effect of bulk FM metals and alloys [3-5] is still widely used for various applications. This is in spite of the fact that the magnitude of the magnetoresistance (MR), i.e., the relative change of the electrical resistivity under the application of an external magnetic field, is usually much smaller for the latter effect. AMR-based sensors are utilized in current-measuring devices [6], electronic compasses (in navigation systems of cars, ships, submarines, and also in GPS locators and mobile phones [7]) and are used as magnetoresistive sensors [8,9] such as position detectors (both linear and angular) and magnetic field detectors.

AMR sensor materials are typically applied in the form of thin films or layers produced by physical deposition methods, mainly evaporation and sputtering. On the other hand, it has long been known [10] that films or layers of the FM elements (Fe, Co and Ni) and many of their FM alloys (with each others or with non-magnetic elements) can be obtained also by electrodeposition (ED). Although there are several reports (see, e.g., Refs. 11-17) on the magnetoresistance (MR) characteristics of FM metals and their homogeneous FM alloys prepared by ED methods, no systematic studies of the AMR have been published on bulk ED FM metals and alloys (the MR investigations on nanoscale ED magnetic heterostructures with GMR effect such as the magnetic/non-magnetic multilayers [18] and the granular magnetic alloys [19] do not fall in this category). This constituted partly our motivation to carry out a MR study on ED Ni-Co alloys throughout the whole composition range.

On the other hand, the magnetoresistance of homogeneous ferromagnets is important also from another point of view. Namely, there is a strong correlation between the GMR magnitude of a FM/NM multilayer and the AMR magnitude of the FM layer material as it will appear from the following observations. With reference to the original note by Snoek [20], it was demonstrated for FM metals and alloys [21-23] that the AMR magnitude exhibits a fairly pronounced maximum as a function of the magnetic moment of the FM material, the maximum being at around 1 $\mu_B$. On the other hand, the GMR magnitude was found [24] to show a maximum as a function of the electron/atom (e/a) number for FM/Cu multilayers where the FM layer is an alloy in the Fe-Co-Ni system; large GMR was observed at e/a values corresponding to alloys with a magnetic moment of about 1 $\mu_B$ as deduced from the well-known Slater-Pauling curves [5]. Miyazaki et al. [25] have displayed these AMR and GMR results also on a ternary alloy diagram directly and large GMR values could, indeed, be identified in the vicinity of alloy compositions exhibiting a large AMR magnitude. Therefore, when looking for FM alloy systems which could yield a possibly large GMR in an FM/NM multilayer, the knowledge of the AMR magnitude may serve as an important first guideline. Sputtered Ni-Co/Cu multilayers have been reported to yield fairly good GMR characteristics [24,26] and this system is easily accessible to the ED preparation technique as well. On the other hand, the reported GMR behaviour of ED Ni-Co/Cu multilayers [18] is still inferior to those of the physically deposited counterparts. Therefore, for the improvement of the GMR of ED Ni-Co/Cu multilayers, a detailed study of the AMR of ED Ni-Co alloys seems to be worthwhile.

It should also be noted in addition that a significant progress has been made recently in the theory of calculating the electrical transport properties of FM metals and alloys, including the residual resistivity and the AMR [27]. For assessing the efficiency of these theoretical approaches and the validity of eventual approximations, reliable experimental data are necessary. However, many of the commonly available data on the electrical transport properties of bulk Ni-Co alloys are rather old, sometimes even contradictory (see, e.g., Fig. 8-21 in Bozorth's book [3]) and not always spanning over a significant composition range.

It was decided, therefore, to carry out a systematic resistivity and MR study of Ni-Co alloys produced by electrodeposition throughout the whole concentration range at and below room-temperature. The investigations were carried out with the FM alloy deposits both on their substrates and also after removing them from the substrates, in the latter case even after a thermal treatment for some samples. The total thickness of the deposited Ni-Co alloy layers was typically 2 μm.

In the Ni-Co system the room-temperature equilibrium phase [28] is a face-centered cubic (fcc) structure up to about 65 at.% Co above which the stable phase is a hexagonal close-packed (hcp) structure, although non-equilibrium fcc Ni-Co alloys with a different microstructure are known to form also in this latter composition range. This means that information on crystal structure and microstructure may also be important when evaluating the magnetoresistance characteristics of Ni-Co alloys. Therefore, X-ray diffraction studies were also carried out on several ED Ni-Co alloys investigated, both before and after annealing.

## 2. Experimental details

*2.1 Sample preparation and characterization*

In order to deposit Ni-Co alloys, two different aqueous electrolytes were first prepared. Each contained the sulphate of one of the two constituent metals only. The composition of the first electrolyte was 0.3 mol/ℓ $Na_2SO_4$, 0.25 mol/ℓ $H_3BO_3$, 0.15 mol/ℓ $H_2NSO_3$ and 0.6 mol/ℓ $NiSO_4$. The last component was substituted in the second one with 0.6 mol/ℓ $CoSO_4$. The pH was set to 3.25 by adding NaOH to both solutions. The choice of this pH value was based on some preliminary experiments to get appropriate deposition conditions. The $Ni^{2+}$- and $Co^{2+}$-containing stock solutions were mixed in appropriate ratios to obtain the electrolytes used for alloy deposition. With this method, we were able to softly tune the composition of the Ni-Co alloys over the entire composition range since there is a strong correlation between the relative ion concentration of cobalt in the electrolyte and the Co-content in the Ni-Co deposits obtained with a constant current density (Fig. 1). The deposit composition was measured with electron probe microanalysis (EPMA) in a JEOL JSM-25 scanning electron microscope. The results in Fig. 1 show that the Ni-Co system exhibits anomalous codeposition which means that although nickel is more noble than cobalt, in all deposits obtained from an electrolyte containing both Co and Ni ions, the Co-content is higher than the relative ion concentration of cobalt in the solution [10].

The Ni-Co alloy samples were deposited on a [100]-oriented, 0.26 mm thick silicon wafer covered with a 5 nm Cr and a 20 nm Cu layer by evaporation. The purpose of the chromium layer was to assure adhesion and the copper layer was used to provide the electrical conductivity of the cathode surface. The surface roughness of the Si/Cr/Cu substrate was investigated with AFM and it showed height fluctuations not larger than 1 nm [29].

The deposition was performed in a tubular cell of 8 mm x 20 mm cross section with an upward looking cathode at the bottom of the cell [30]. Direct-current (d.c.) deposition was carried out by using either $-31.3$ mA/cm$^2$ or $-18.8$ mA/cm$^2$ current densities.

For studies of the composition dependence of the physical properties of Ni-Co alloys, the deposition time was chosen to get the same thickness (about 2 μm) for all samples, assuming 100% current efficiency. It was established from profilometric measurements carried out on several samples that the actual current efficiency was 96%.

Some Ni-Co alloy samples removed from their Si substrates were subjected to an annealing carried out at 300ºC for one hour in $H_2$ atmosphere.

X-ray diffraction (XRD) was used to investigate the crystal structure, the texture and the microstructure of some selected Ni-Co alloy deposits with the help of a Philips X'pert powder diffractometer in the $\Theta$-2$\Theta$ geometry with Cu-K$_\alpha$ radiation. The lattice parameter and the average crystallite size of the alloys were calculated from the position and the full width at half maximum (FWHM), respectively, of the XRD peaks. These data were determined by fitting Pearson-VII curves on the background-corrected and smoothed XRD diffractograms.

*2.2 Measurement of electrical transport properties*

The electrical transport measurements were carried out in the field-in-plane/current-in-plane geometry by applying a d.c. current in different four-point-in-line probes with various distances of the current and potential contact points depending on the particular sample geometry and parameter measured.

The magnetoresistance (MR) was measured as a function of the external magnetic field (H) up to 8 kOe. The MR ratio was defined with the formula

$$MR(H) = \frac{\Delta\rho}{\rho_0} = \frac{\Delta R(H)}{R_0} = \frac{R(H) - R_0}{R_0}, \qquad (1)$$

where Δρ and ΔR(H) are the change of the sample resistivity and resistance due to the magnetic field, respectively, whereas $\rho_0$ and $R_0$ are the zero-field resistivity and resistance of the sample, respectively, and R(H) is the resistance in an external magnetic field H.

The room-temperature MR data were first measured for the whole alloy deposit (8 mm x 20 mm) in the as-prepared state still on the Si/Cr/Cu substrate. After peeling off the deposits mechanically from the wafer, 2-3 mm wide strips were cut from the samples and then attached to a scotch tape to carry out measurements in another probe. Both the longitudinal (LMR, magnetic field parallel to the current) and the transverse (TMR, field perpendicular to the current) components of the magnetoresistance were measured for each sample.

The zero-field resistivity $\rho_0$ was determined at room-temperature with a probe calibrated with copper foils of known thickness and having the same lateral dimensions as the alloy samples.

The low-temperature resistivity and MR measurements were carried out in a Leybold LTC60-type closed-circuit He-cryostat in which 13 K could be reached and held stable. The temperature was measured with a LakeShore Cryotronics DRC 81C-type semiconductor thermometer.

For the residual resistivity measurements, the cryostat was first cooled down to 13 K, then the whole system was heated up to room-temperature with a constant heating power while the resistance of the sample was continuously measured as a function of temperature.

## 3. Results and discussion
*3.1 Structural characterization*

In the as-deposited state (either on the Si/Cr/Cu substrate or after removing the substrate), all Ni-Co alloys which were investigated by XRD exhibited a very pronounced texture in that, besides a large-intensity Bragg reflection, several small peaks could only be revealed with intensities amounting to at most a few percent of the main peak. All the observed XRD peak positions could be identified with the expected reflections of an fcc phase. By indexing to an fcc lattice, the intensity ratio of the (220) and the (111) peaks ($I_{220}/I_{111}$) was about 35 for all ED Ni-Co alloys investigated by XRD in the as-deposited state, indicating a very pronounced (110) texture. After annealing, however, this ratio drastically changed, it became nearly unity, indicating a loss of the original preferred texture to a large degree.

Because of the uncertainty of the parameters of the small peaks of the diffractograms, only the data determined from the sufficiently intense (220) peak were used for both the unannealed and annealed samples to derive reliable lattice parameter values. The measured lattice parameters are plotted in Fig. 2 along with the data found in the literature for the Ni-Co alloy system. The results show a good agreement both with Vegard's law by using the lattice parameters of fcc-Ni and fcc-Co [31] and with previous data on metallurgically processed bulk Ni-Co alloys [32,33].

Since above about 65 at.% Co the equilibrium phase of Ni-Co alloys is the hcp structure [28], we have checked also for the possibility of hcp phase formation in our samples. It turned out that whereas the position of the main fcc peak (220) strongly coincides with the expected position of a hcp reflection when scaled with Vegard's law between hcp-Co [31] and hcp-Ni [34], the other XRD peak positions observed for the present ED Ni-Co alloys (even after annealing) do not correspond to any of the expected hcp reflections in the composition range of the equilibrium hcp Ni-Co alloys. Therefore, the ED Ni-Co alloys investigated in this work probably all exhibit a predominating fcc structure, although the presence of some hcp fraction with a very strong texture cannot be excluded either. Selected-area transmission electron microscopy studies would be necessary to get firm evidence about the absence or presence of a hcp phase in the samples with Co-contents above 65 at.%.

As to the microstructure of the ED Ni-Co alloys, the Scherrer equation [35] was used to determine the maximum of the average crystallite size $D$. As it can be seen in Fig. 3, D is around 10 nm throughout the entire concentration range, i.e., the as-deposited ED Ni-Co alloys exhibit a nanocrystalline structure. No systematic difference between the grain sizes of samples deposited at the two current density values applied could be observed.

The nanocrystalline state is preserved also after the annealing applied although the average crystallite size increased by about a factor of 3 to 5 (Fig. 3). According to the results obtained, the average crystallite size monotonously decreases with increasing cobalt content in the annealed state.

*3.2 Room-temperature electrical transport properties*
3.2.1 Zero-field electrical resistivity

The room-temperature zero-field electrical resistivity ($\rho_0$) of the ED Ni-Co alloys was first measured before annealing in the as-deposited state of the samples, i.e., while still being on their Si/Cr/Cu substrates and before putting them in a magnetic field. For annealing, the Ni-Co alloy deposits were removed from their substrates. After the annealing was carried out and the magnetoresistance was measured, the zero-field resistivity of the annealed state was calculated from the voltage reading at $H = 0$ of the magnetoresistance measurement.

Figure 4 shows the zero-field resistivity ($\rho_0$) data in various states of the ED Ni-Co alloys in comparison with relevant literature data [21,36-39]. The $\rho_0$ values for the unannealed state measured with the alloy samples on their substrate (Fig. 4a) were practically the same for the samples deposited at either current densities. The same figure indicates that after annealing the stripped alloy samples, the resistivity significantly decreased. According to our recent work [39], the room-temperature resistivity of the Cr(5nm)/Cu(20nm) bilayer on the Si wafer is 6.2 μΩ·cm. With this data, we can estimate that the shunting effect of this metallic bilayer on the measured resistivities of the Ni-Co alloys having a thickness of 2 μm is less than 1 % of the measured values for $\rho_0$ values not exceeding 10 μΩ·cm; however, even for $\rho_0 = 20$ μΩ·cm, this correction is about 2.5 % only. By considering that during stripping the Cr/Cu underlayers may partly be retained on the substrate, the actual shunting effect is certainly smaller, i.e., the above specified correction values are probably overestimated.

It should be noted that when removing the samples from the Si/Cr/Cu substrate, the electrical resistivity slightly increased (by about 20 %). This can be partly due to the mechanical deformation of the alloy deposits during the mechanical stripping process. However, this deformation effect is completely removed by the annealing procedure.

The room-temperature resistivity data for the annealed ED Ni-Co alloys compare well (Fig. 4b) to results reported for a few compositions on metallurgically processed Ni-Co alloys [21,36,37]. The data on the annealed ED Ni-Co alloys also smoothly join the room-temperature resistivity values reported for the pure fcc-Ni [38] and fcc-Co [39] metals.

The resistivity change upon annealing can be explained by taking into account the influence of grain boundaries on the electrical transport. Grain boundaries represent lattice imperfections and the scattering of electrons on them reduces the flow of electricity, i.e., their presence leads to a grain boundary contribution to the resistivity [40]. It was reported formerly, indeed, that the electrical resistivity can be significantly larger in nanocrystalline ED Ni [41-44] and Co [44,45] in comparison with the resistivity of well-annealed, coarse-grained state of these metals. For rationalizing the electrical resistivity differences in Fig. 4a, we need to make use of an estimate on the relative fraction of grain-boundary atoms in the nanocrystalline state as a function of the grain size. According to the work of Siegel [46], by assuming a grain boundary width of 1 nm, the relative fraction of atoms in the grain boundaries of our nanocrystalline metals is about 25 % at $D = 10$ nm, about 10 % at $D = 30$ nm and about 5 % at $D = 50$ nm.

On the basis of these data, we can say that the very large resistivities for the unannealed ED Ni-Co alloys (Fig. 4a) are due to the 10-nm grain size of the as-deposited alloys.

Along the same line, the significant reduction of the resistivity of ED Ni-Co alloys upon annealing can be ascribed to an increase of the grain size during heat treatment. With reference to the above estimate on the grain boundaries, the reduction of the volume fraction of the grain-boundary atoms from about 25 % to the level of 5 to 10 % caused a resistivity reduction by about 50 % throughout the whole composition range. By considering the differences in the grain size between the Ni-rich and Co-rich ends of the Ni-Co alloy compositions, we can say that at the Ni-rich end the grain-boundary contribution to the measured resistivity is smaller than at the Co-rich end.

3.2.2 Magnetoresistance

Figure 5a displays the measured $MR(H)$ data for an annealed Ni-Co alloy which was chosen to show a behaviour very similar to that of an isotropic FM metal. The positive part of the curve (full symbols) represents the LMR data and the negative part (open symbols) the TMR data. This is the typical behaviour for most FM metals and alloys [3-5] which is explained in Fig. 5b by using the data of Fig. 5a. According to this (i) both MR components vary rapidly (LMR: increases; TMR: decreases) when increasing the magnetic field from $H = 0$ until technical saturation ($H = H_s$), (ii) the LMR component always remains positive and the TMR component negative in the whole range of magnetic fields investigated, (iii) for magnetic fields beyond saturation ($H > H_s$), the MR data exhibit an approximately linear decrease with increasing magnetic field and (iv) for the saturation MR values obtained by extrapolation to $H = 0$, the relations $LMR_s > 0$ and $TMR_s < 0$ hold.

The rapid variation of the magnetoresistance for small magnetic fields corresponds to the magnetization process by domain wall displacement whereas the approach to saturation around $H_s$ to the gradual alignment of the magnetization of the rest of the domains until technical saturation is reached. In magnetic fields higher than the saturation field, at $T = 0$ K the magnetic moments of the individual atoms point at the direction parallel to the applied external magnetic field. However, at a finite temperature, thermal excitations make these magnetic moments fluctuate around their equilibrium orientation. As the magnetic field is increased, these fluctuations are continuously reduced because the thermal excitation energy is not sufficient to randomize the magnetic moments with respect to the direction of the applied field as much as it was in a lower field. This effect is called paraprocess due to which the resistivity decreases because the scattering of the current-carrying (conduction) electrons on the less and less fluctuating moments is reduced. This reduced scattering results in a practically linear resistivity decrease with increasing magnetic field. Since the resistivity decrease is the same for any mutual orientation of the measuring current and the magnetization, both the LMR and TMR components exhibit the same decreasing slope in the paraprocess region. By applying a linear fit to the saturation region ($H > H_s$) of the $MR(H)$ data (Fig. 5b), both the LMR and the TMR component can be extrapolated to zero field which procedure yields the saturation values $LMR_s$ and $TMR_s$. The difference between the two components defines the magnitude of the anisotropic magnetoresistance: $AMR = LMR_s - TMR_s$.

The physical origin for the difference in the $LMR(H)$ and $TMR(H)$ curves shown in Fig. 5 comes from the fact that for a ferromagnetic sample magnetized until saturation (what implies that it is in a single-domain state), the resistivity depends only on the angle between the magnetization (the direction of which is determined by the magnetic field) and the measuring current [5]:

$$\frac{\Delta\rho(H)}{\rho_{av}} = \frac{\Delta\rho}{\rho_{av}}\left(\cos^2\Theta - \frac{1}{3}\right) \qquad (2)$$

where $\rho_{av} = (\rho_L + 2\rho_T)/3$ is the average (isotropic) resistivity and $\Theta$ is the angle between the applied magnetic field and the magnetization of the sample. $\Theta = 0$ stands for the LMR component and $\Theta = \pi/2$ stands for the TMR component. This yields the ratio $TMR/LMR = -1/2$ for an isotropically demagnetized material. We call a demagnetized state (i.e., $M = 0$) isotropic when the magnetization orientations of the domains are random. A deviation of the ratio $TMR_s/LMR_s$ from -1/2 may occur if, due to a magnetic anisotropy, the magnetization orientation distribution in the remnant state of the FM specimen is not random.

Before presenting the results of magnetoresistance measurements on the ED Ni-Co alloys, it is noted that the MR behaviour was found to vary both with alloy composition and with the state of the samples (on substrate or without substrate, annealed or unannealed). This multitude of MR characteristics necessitates the subdivision of the presentation of the experimental results which are primarily arranged according to the sample states from the as-deposited state on the substrate to annealed state without substrate.

*3.2.2.1 Magnetoresistance results in the as-deposited state on substrate*

In the as-deposited state on substrate, the $MR(H)$ curves were found to be qualitatively different depending on the alloy composition.

(i) For those as-deposited ED Ni-Co alloy samples on substrate in which *the Co-content was higher than about 50 at.%*, the *MR(H)* curves were found to be qualitatively similar to that shown in Fig. 5a. This corresponds to the usual behaviour observed for most bulk ferromagnetic alloys. In this composition range the $TMR_S/LMR_S$ ratio was found to be about -1 and did not show any systematic dependence on alloy composition. The deviation of this ratio from -1/2 indicates the presence of some magnetic anisotropy in these alloys.

(ii) For those as-deposited ED Ni-Co alloys in which *the Co-content was smaller than about 50 at.%*, a surprising observation was made Namely, whereas the LMR component exhibited the usual behaviour also here (*LMR* > 0 for all fields), the TMR component showed, instead of the usual decrease, an increase for small magnetic fields until saturation. This initial increase was sometimes very small so that the *TMR(H)* values became negative for sufficiently high magnetic fields (Fig. 6a). However, in some cases, the initial increase of TMR was so significant that the TMR component remained positive in the whole range of magnetic fields applied (Fig. 6b). Beyond technical saturation, the usual linear decrease of the *MR(H)* data could be observed also for this Co-content range. In any case, the extrapolated $TMR_S$ values were definitely positive for these samples. The magnitude of the positive upturn of the TMR component (and, thus, the value of $TMR_S$) was especially large for alloy compositions between 10 to 30 at.% Co. Due to the $TMR_S > 0$ values, the $TMR_S/LMR_S$ ratios were also positive here.

The results presented in Fig. 6 underpin the importance of measuring the full dependence of the magnetoresistance as a function of the magnetic field since, depending on the magnetic field value, in extreme cases the TMR component can be either positive or negative. If the magnetoresistance is measured at a single field value only, one cannot properly understand the behaviour of the individual MR components and this may completely mislead the physical interpretation of experimental results in some cases.

The dependence of the *AMR* on the Co-concentration of ED Ni-Co alloys is shown in Fig. 7a for the as-deposited state on substrate by the open circles.

*3.2.2.2 Magnetoresistance results on unannealed samples after removal from their substrates*

For these studies, the ED Ni-Co alloy samples were detached from the Si/Cr/Cu substrates and were put on a scotch tape after which the *MR(H)* curves were measured again. It was observed that for all alloys investigated after removal from the substrate, the behaviour of the TMR component changed and became similar to that of Fig. 5 in the whole composition range: immediate decrease for small magnetic fields and yielding a negative $TMR_S$ value. This is an important indication that the origin of the positive $TMR_S$ values for Co-contents below about 50 at.% Co as described in the preceding secrtion should lie in the interaction between the Ni-Co alloy film and the substrate and this interaction should be of magnetostrictive origin. However, further studies are required to properly understand the origin of the positive TMR values in these ED Ni-Co alloy films and this work is in progress.

In agreement with the slight increase of the zero-field resistivity values after removal from the substrate (see Section 3.2.1), the *AMR* values for the unannealed Ni-Co alloy samples removed from their substrate (open triangles in Fig. 7a) were somewhat smaller than the corresponding data measured for the samples in the as-deposited state on the substrate (open circles in Fig. 7a).

*3.2.2.3 Magnetoresistance results on annealed samples without their substrates*

The *MR(H)* curves were measured also after annealing the samples without their substrates. For all annealed samples, the $TMR_S/LMR_S$ ratio was found to be very close to -1/2, indicating that an important effect of the annealing is the removal of any magnetic anisotropy from the alloys, in accord with the strong reduction of the degree of texture upon annealing (see Section 3.1). The *AMR* values obtained after annealing are also presented in Fig. 7a (full squares). They showed significant changes (increase) for high Co-contents only.

In Fig. 7b, the room-temperature AMR data obtained for the annealed ED Ni-Co alloys are displayed in comparison with literature data on metallurgically processed Ni-Co alloys [21,37] and on Ni-Co alloy films [47]. A fairly good agreement of data on alloys prepared by various methods can be observed throughout the whole concentration range. Similarly to the room-temperature zero-field resistivity (Fig. 4b), also the *AMR* data smoothly join the values reported for the pure Ni and Co metals.

*3.3 Low-temperature electrical transport properties*
3.3.1 Temperature dependence of the electrical resistivity

The temperature dependence of the zero-field resistivity $\rho_0$ was measured between 13 K and 300 K. The variation of $\rho_0$ with temperature can be seen in Fig. 8 for a selected ED Ni-Co alloy. The overall shape of the evolution of $\rho_0$ with temperature was found to be very similar for all samples investigated. For comparison, we have included in Fig. 8 also the only available dataset [21] reported for the temperature dependence of $\rho_0$ for a metallurgically processed Ni-Co alloy and the corresponding data for well-annealed pure fcc-Ni [48,49].

In ferromagnetic metals, the electrical resistivity below the Curie point arises from two major contributions [40,50]: electron-phonon and electron-magnon scattering. The temperature dependence of the electrical resistivity due to the electron-phonon scattering is described by the Bloch-Grüneisen formula [40,50] which depends only on the ratio $T/\Theta_D$ where $\Theta_D$ is the Debye temperature. In the low-temperature limit ($T \ll \Theta_D$), the Bloch-Grüneisen formula simplifies to the $T^5$-law of the temperature dependence of the resistivity of metals [40,50] which was indeed found to be valid experimentally for numerous metals. According to the low-temperature specific heat measurements on fcc Ni-Co alloys by Caudron et al. [51], the Debye temperature changes only slightly with Co-content (from 450 K to 433 K as the Co-content increases from 0 to 65 at.%). It should be noted that $\Theta_D$ has a very close value even for hcp-Co metal (445 K [38]). All this means that the temperature dependence of the resistivity due to the electron-phonon scattering should be very similar for Ni-Co alloys in the whole concentration range for $T$ below about 50 K. In this temperature range, the electron-magnon scattering is still weak due to the high Curie temperatures ($T_c$ = 631 K for Ni and the Curie point continuously increases with the addition of Co [28]). With increasing temperature, due to the increasing excitation of spin-waves, the electron-magnon contribution to the resistivity will be progressively larger.

In our case, since we are well below the Curie-temperature, we see a good agreement between our results and the previously reported temperature evolution of resistivity. This means that the fine-grained structure of the ED Ni-Co alloys has no measurable effect on the Debye-temperature.

3.3.2 Residual resistivity

On the basis of the measured temperature dependence of the resistivity in the range from 13 K to 300 K, the residual resistivities ($\rho_{res}$) were determined from the data for the lowest temperatures investigated. The $\rho_{res}$ results for the ED Ni-Co alloys are displayed in Fig. 9 (full triangles).

As for the room-temperature results, also our residual resistivity data compare well with values reported for metallurgically processed Ni-Co alloys. On the other hand, it should be noted that the relatively large scatter of the experimental data in Fig. 9 may come, in addition to measurement uncertainties, from the different impurity content and/or microstructure of the various Ni-Co alloys. Both these features may lead to various lattice imperfections which can have a significant influence on the residual resistivity. We should also keep in mind that the different processing methods for the preparation of the Ni-Co alloy samples may have resulted in various degrees of chemical short-range order, i.e., not necessarily all samples investigated exhibited a chemically perfectly disordered state (random solid solution). Unfortunately, no information is available on the degree of chemical disorder in any of the Ni-Co alloys for which experimental data have been reported. On the other hand, the nanocrystalline state with the grain sizes reported in Section 3.1 for the ED Ni-Co alloys in the annealed state (30 to 50 nm) apparently does not give a significant contribution to the residual resistivity.

We should also make a remark concerning the evolution of $\rho_{res}$ in the Ni-Co alloy system. For a disordered (random) binary alloy $A_{1-x}B_x$, the electrical resistivity contribution due to chemical disorder (as reflected by $\rho_{res}$) usually follows Nordheim's rule [50] according to which $\rho_{res} \propto x(1-x)$. This formula yields a parabolic behaviour which is well obeyed by several binary (mostly copper-based) alloys [50]. The Nordheim's rule is more or less properly followed by the residual resistivity data for the Ni-Co alloys (Fig. 9) although there seems to be a tendency in that the maximum is shifted to Co-contents around 20 to 30 at.%. The reason for this deviation may partly come from the fact that for alloys containing transition metals and especially FM alloy systems, Nordheim's rule is not necessarily valid. An eventual deviation from perfect chemical disorder may also contribute to a shift of the resistivity peak position from the equiatomic composition.

There is only one theoretical calculation for the residual resistivity of fcc Ni-Co alloys which was reported by Banhart et al. [27b] and their data are also displayed in Fig. 9 by the crosses (x). A comparison with the experimental data reveals that the theoretical values qualitatively very well account for the composition dependence of the residual resistivity in the fcc Ni-Co alloy system, although the calculated values are smaller in magnitude by about a factor of 5. One should keep in mind that the parameter-free calculations were carried out for a structurally perfect crystal with complete chemical disorder whereas the experimentally investigated alloy samples are certainly less perfect and may have a different degree of chemical disorder. This may partly explain the larger experimental values and recent improvements in theoretical calculations are also expected to yield a better quantitative agreement with experiments when applied to this system.

3.3.3 Magnetoresistance

The *MR*(*H*) curves were measured at 13 K in both the longitudinal and the transverse configurations for some selected ED Ni-Co alloy samples without substrate either in the unannealed or annealed state. For the ED Ni-Co alloys with a Co-content not exceeding about 70 at.%, the field evolution of the LMR and TMR components was found to be very similar to the typical room-temperature curves which were shown in Fig. 5a. From the linear sections for $H > H_s$, the saturation *LMR* and *TMR* values were established from which the *AMR* values were deduced according to eq. (2). The ratio of the $LMR_s$ and $TMR_s$ was -1/2 in this compositon range, which is the case for isotropic alloys. However, for the Ni-Co alloys having a Co-content above about 70 at.%, the *MR*(*H*) curves did not completely reach saturation for both components even at the highest applied field (8 kOe); therefore, the determination of the *AMR* value was not so straightforward and, also, the error might be larger for this case. The larger saturation fields indicate an increased magnetic anisotropy of the Ni-Co alloys for high Co-contents at low temperatures.

The *AMR* values determined at $T = 13$ K for the ED Ni-Co alloys are presented in Fig. 10 and compare well with available low-temperature literature data for metallurgically processed Ni-Co alloys which are also displayed. It seems that the nanocrystalline state of the ED Ni-Co alloys does not influence significantly the low-temperature *AMR* values. Furthermore, although there is a fairly large scatter of the experimental data from various reports, a clear maximum in the vicinity of about 20 to 30 at.% Co can be revealed. Nevertheless, the low-temperature *AMR* values are much higher than the data obtained at room temperature (Fig. 7). This indicates that the temperature variation of the *AMR* is dominated by the strong reduction of the zero-field resistivity with decreasing temperature.

Banhart et al. [27b] reported a theoretical calculation also for the *AMR* of Ni-Co alloys at $T = 0$ K and their data are displayed in Fig. 10 by the crosses (×). As for the resistivity, the composition evolution of the theoretical *AMR* values also properly reproduces the trend exhibited by the low-temperature experimental data. On the other hand, Fig. 10 also shows that the calculated values are larger by about a factor of 2 in comparison with the experimental data. This partly comes from the too low calculated residual resistivities (Fig. 9) but certainly there is still a discrepancy also with the value of the calculated resistivity difference between the longitudinal and transverse MR components. The situation will certainly improve also in the case of the magnetoresistance by the application of recent developments in theoretical calculations for the Ni-Co ally system.

## 4. Summary

In the present work, the electrical transport properties (zero-field resistivity, temperature dependence of the resistivity, residual resistivity and the AMR) were investigated for Ni-Co alloys. For this purpose, Ni-Co alloy layers having a thickness of about 2 μm were prepared by electrodeposition on Si wafers with evaporated Cr and Cu underlayers in the whole composition range. The Ni-Co alloy deposits were investigated first in the as-deposited state on the substrates and then, by mechanically stripping them from the substrates, as self-supporting layers both without and with annealing.

It was established by XRD that the ED Ni alloy deposits exhibited an fcc phase. A strongly textured fcc structure was found in the as-deposited state with an average grain size of about 10 nm. Upon annealing, whereas the crystal structure was retained, the grain size increased by a factor of 3 to 5, depending on alloy composition. The room-temperature zero-field resistivity was found to decrease strongly by annealing due to the increased grain size.

By measuring the temperature dependence of the resistivity down to 13 K, the residual resistivities of the annealed ED Ni-Co alloys were determined which were in good agreement with previously reported data on metallurgically processed Ni-Co alloys. The residual resistivity exhibits a maximum as a function of composition for the Ni-Co alloy system, corresponding to Nordheim's rule, although the maximum position is shifted from the expected equiatomic composition to about 20 to 30 at.% Co.

The *AMR* values measured at room temperature and 13 K on the substrate-free ED Ni-Co alloys either in unannealed or annealed state were in relatively good agreement with reported data on bulk Ni-Co alloys prepared by metallurgical means. The *AMR* values also exhibit a maximum in the same composition range as the residual resistivity in this alloy system.

The only theoretical calculation reported previously in the literature for $T = 0$ K on the corresponding electrical transport parameters of fcc Ni-Co alloys [27b] qualitatively very well described the experimental composition dependence of both the residual resistivity and the low-temperature AMR data. The still existing quantitative differences between theory and experiment may partly be ascribed to the fact that the parameter-free calculations were carried out on idealized materials (structural perfectness and complete chemical disorder) whereas there are structural defects and a possibly different degree of chemical disorder in the experimentally investigated samples. On the other hand, recent progress in theoretical calculations of the electrical transport parameters of ferromagnetic alloys gives promise that future calculations when applied to such systems will yield better agreement with experimental data even quantitatively.

**Acknowledgements**

This work was supported by the Hugarian Scientific Research Fund through grant OTKA K 75008. The closed-cycle He-cryostat was kindly donated by the Humboldt Foundation. The authors thank I. Balogh for performing the annealing treatments and A. Csik (Institute for Nuclear Research, HAS) for the profilometric measurements. The authors also acknowledge G. Molnár and A. Tóth (Research Institute for Technical Physics and Materials Science, HAS) for preparing the evaporated underlayers and providing access to the SEM analytical facility, respectively.

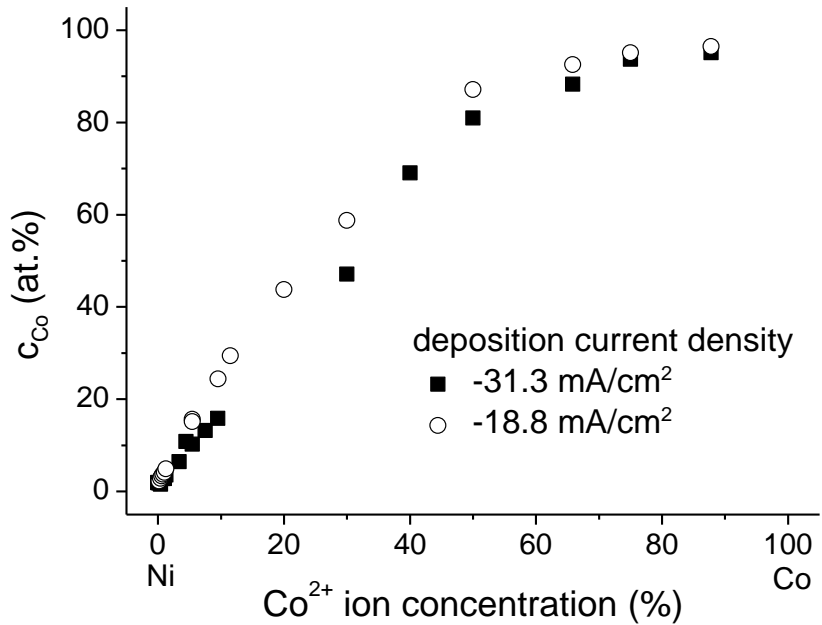

*Fig. 1*  Co-content $c_{Co}$ in the ED Ni-Co alloy deposits for two different current densities as a function of the relative $Co^{2+}$ ion concentration in the solution, the latter quantity defined as $100 \times c(Co^{2+})/[c(Co^{2+}) + c(Ni^{2+})]$.

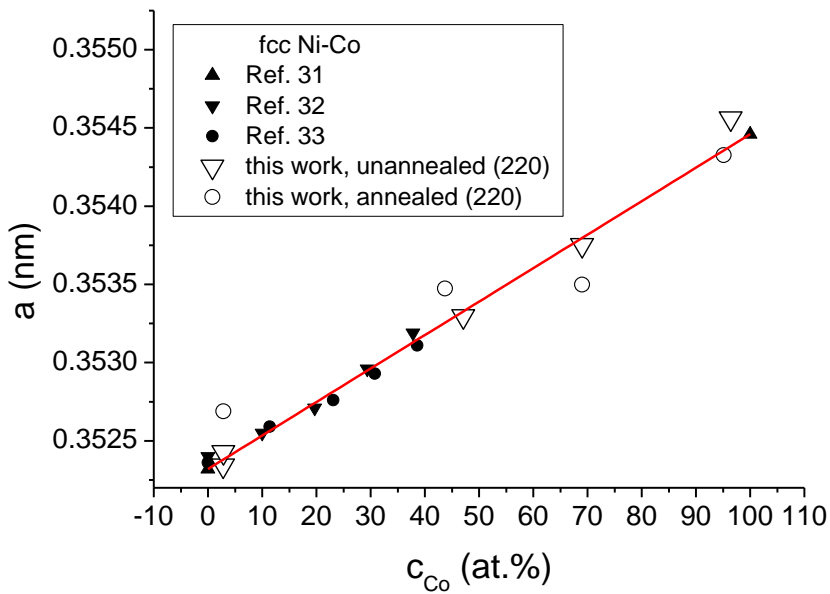

*Fig. 2*  Lattice parameter *a* of ED fcc Ni-Co alloys as a function of the deposit Co-content (open symbols, estimated error: ±5 %). For comparison, literature data (full symbols) for fcc-Ni and fcc-Co metals as well as for metallurgically processed fcc Ni-Co alloys are also displayed. The straight line represents Vegard's law.

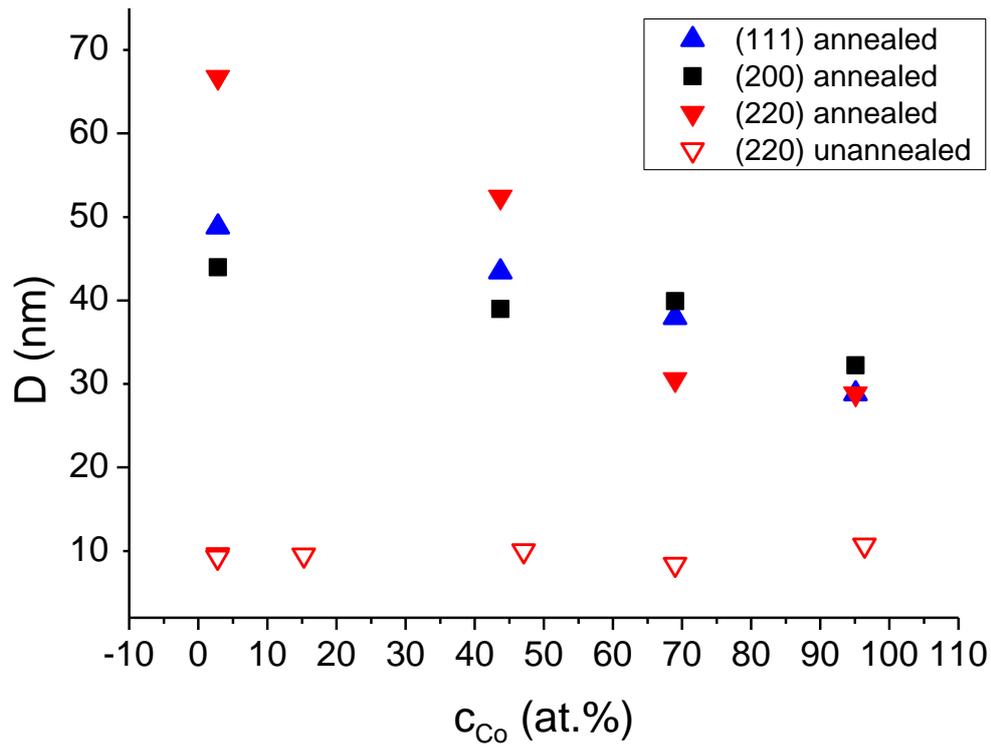

*Fig. 3* Average crystallite size *D* of ED Ni-Co alloys as a function of the deposit Co-content, calculated from different XRD reflections (estimated error: ±5 %). Open symbols represent the data of the samples obtained from XRD in the as-deposited state (before annealing). Full symbols show the data for annealed samples.

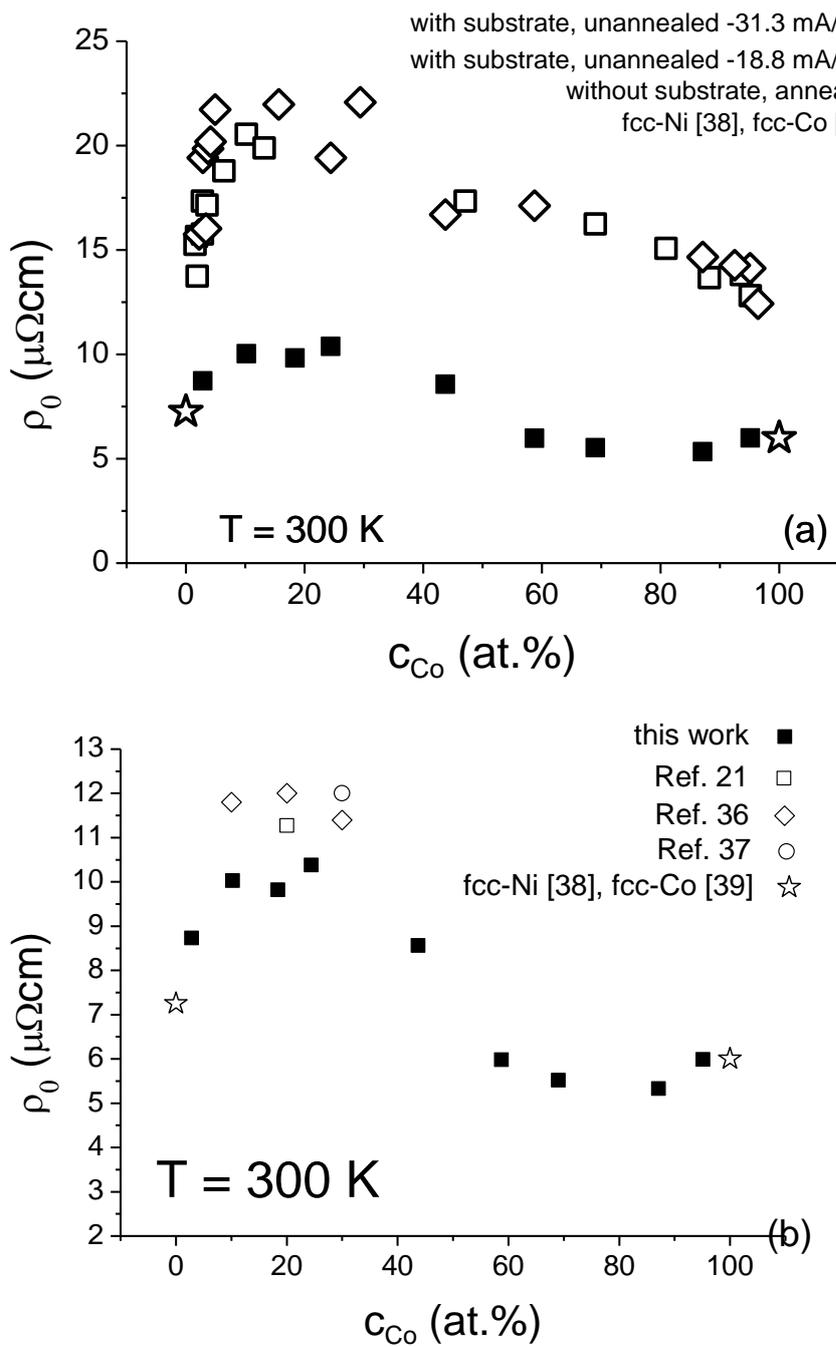

*Fig. 4* Room-temperature electrical resistivity $\rho_0$ of ED Ni-Co alloys as a function of the Co-concentration. (a): present data obtained on samples in the as-deposited state (open ◇ and □ symbols) and after annealing (full ■ symbols); (b) present data obtained on annealed samples (full ■ symbols) in comparison with some literature data on metallurgically processed Ni-Co alloys as indicated in the legend. The stars indicate the resistivity data for the pure fcc metals [38,39].

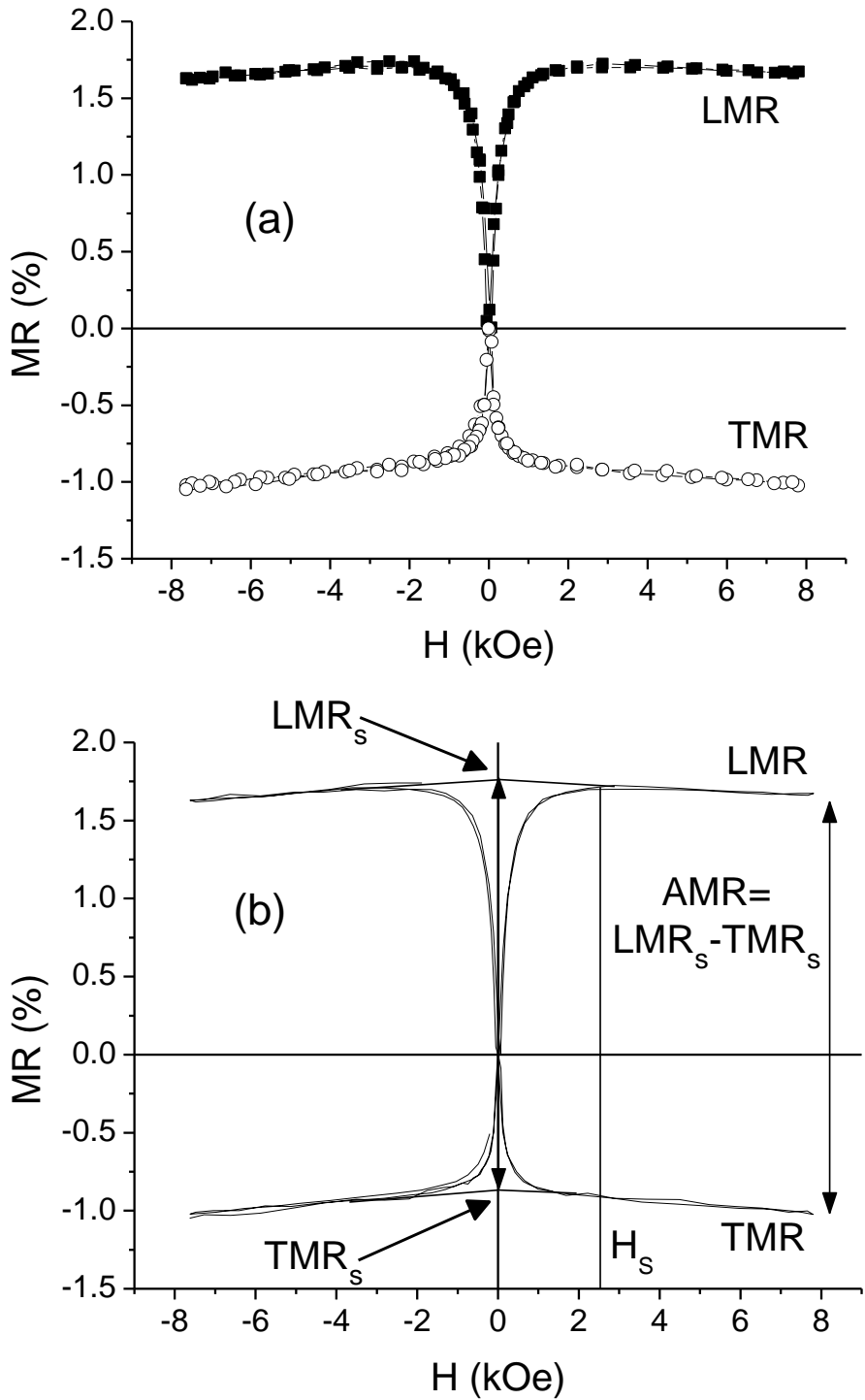

*Fig. 5* (a) Longitudinal (full symbols, LMR) and transverse (open symbols, TMR) magnetoresistance data for an ED $Ni_{97.2}Co_{2.8}$ alloy without substrate and after annealing. The inset shows the relative orientations of the measuring current and the external magnetic field for the LMR and TMR components. (b) The definition of the saturation values $LMR_s$ and $TMR_s$ obtained by linear extrapolation to $H = 0$ from the data of (a) in the magnetic field region above saturation ($H > H_s$).

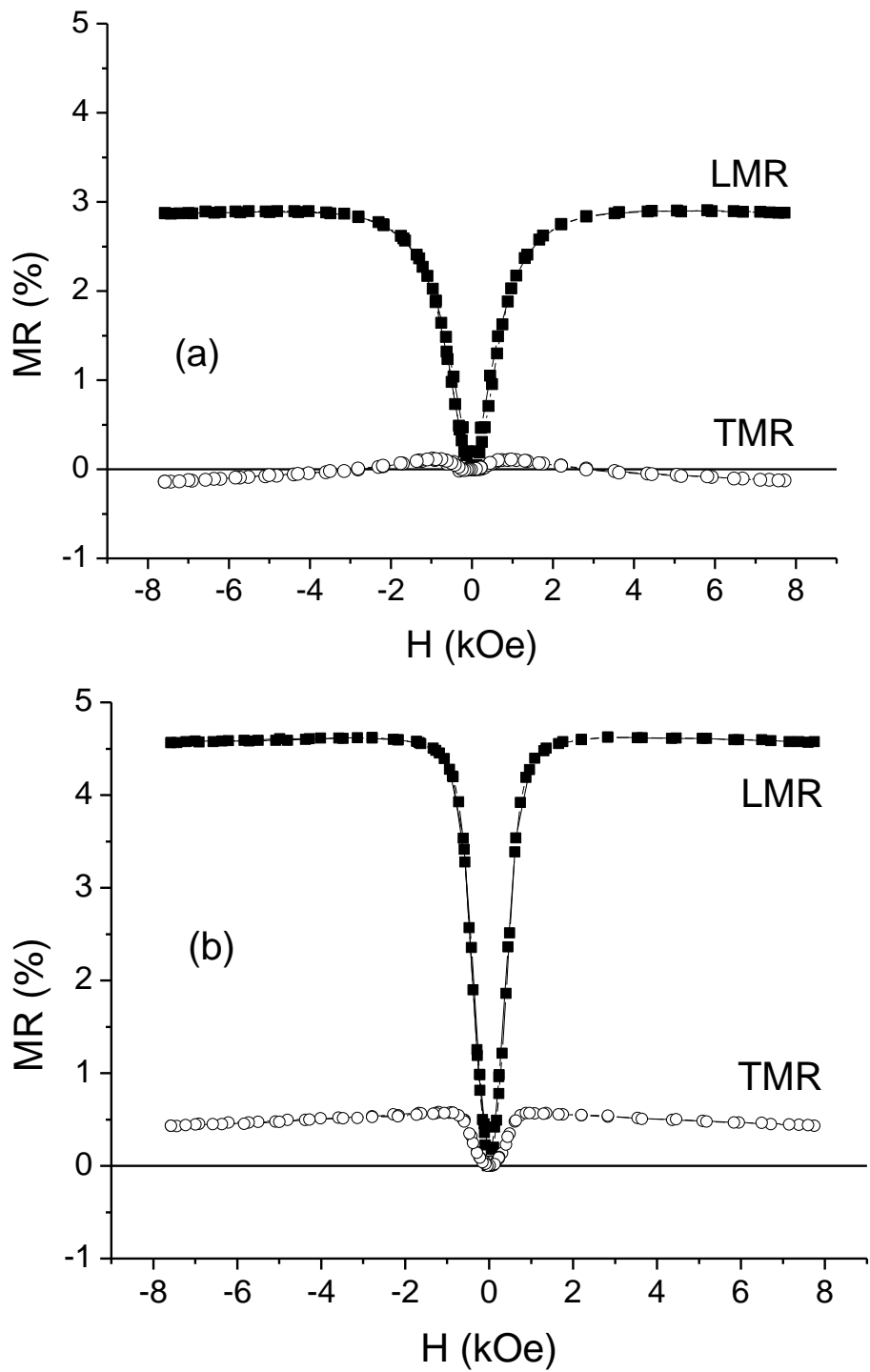

*Fig. 6* Longitudinal (full symbols, LMR) and transverse (open symbols, TMR) magnetoresistance data for two ED Ni-Co alloys in the as-deposited state (on substrate, unannealed). (a) ED $Ni_{97.2}Co_{2.8}$ alloy and (b) an ED $Ni_{72.1}Co_{27.9}$ alloy. In both cases, the TMR component starts to increase for small *H* values before reaching saturation beyond which the *MR(H)* data exhibit a linear decrease with increasing magnetic field. The two cases differ in that the *MR*(*H* = 8 kOe) value is negative in (a) and positive in (b).

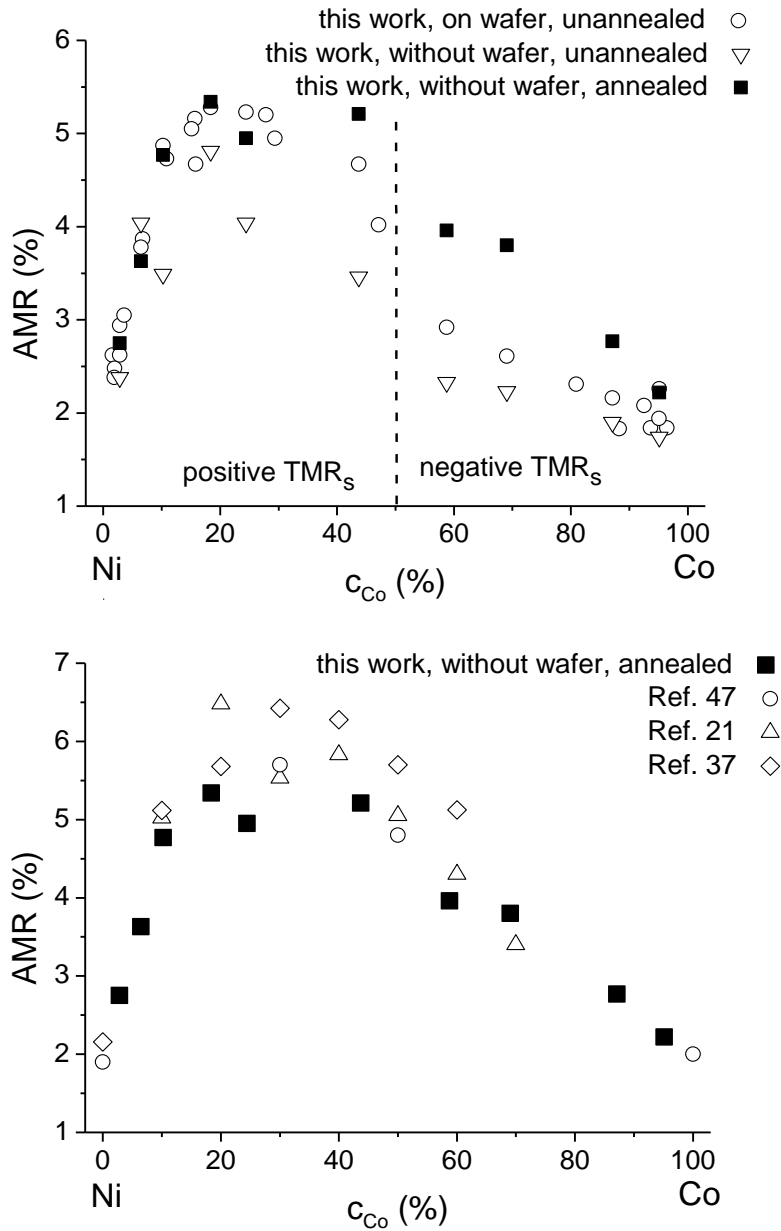

*Fig. 7* (a) Room-temperature *AMR* data for the ED Ni-Co alloys as a function of the Co-content. In the unannealed state (open circles) with the samples still on the substrate, there is a change in the sign of the $TMR_s$ values at about 50 at.% Co-concentration, indicated by the dashed vertical line. The open triangles and full squares show the *AMR* data before and after annealing, respectively, for the samples which were removed from the Si/Cr/Cu substrates; (b) Room-temperature *AMR* values of the annealed ED Ni-Co alloys (full squares) as a function of the Co-content, in comparison with literature data on metallurgically prepared Ni-Co alloys [21,36] and evaporated Ni-Co alloy films [47] (note that the Co value is for the hcp phase).

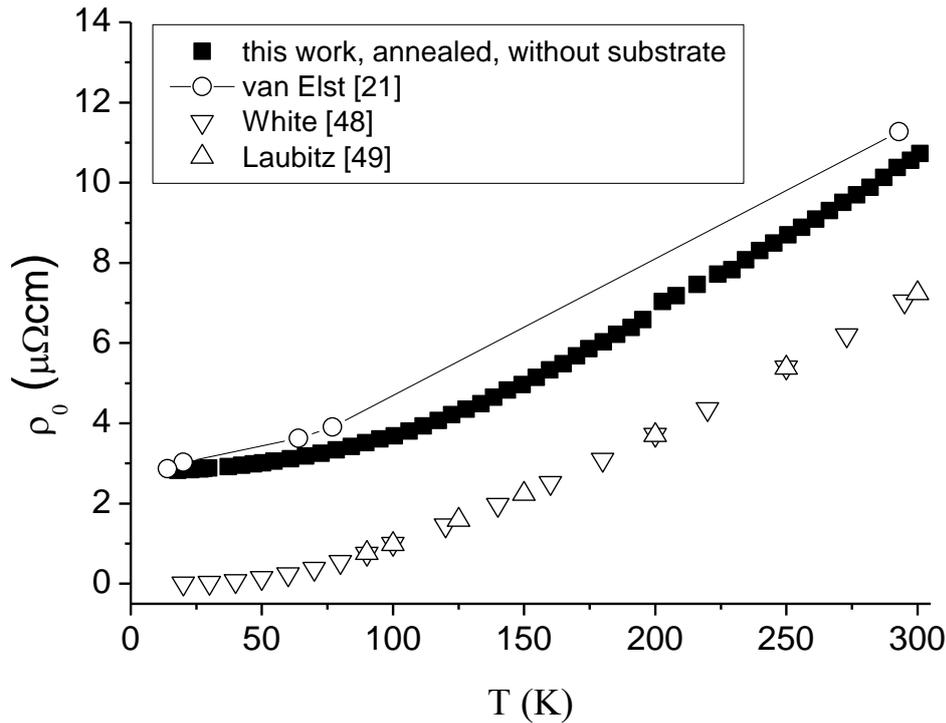

*Fig. 8* Resistivity data (■) of a selected ED Ni-Co alloy ($Ni_{75.6}Co_{24.4}$, annealed, without the Si/Cr/Cu substrate) as a function of the temperature. The open symbols refer to the literature data on a metallurgically processed $Ni_{80}Co_{20}$ alloy (○) [21] and the resistivity data of well-annealed pure nickel ($\nabla$, $\Delta$) [48,49].

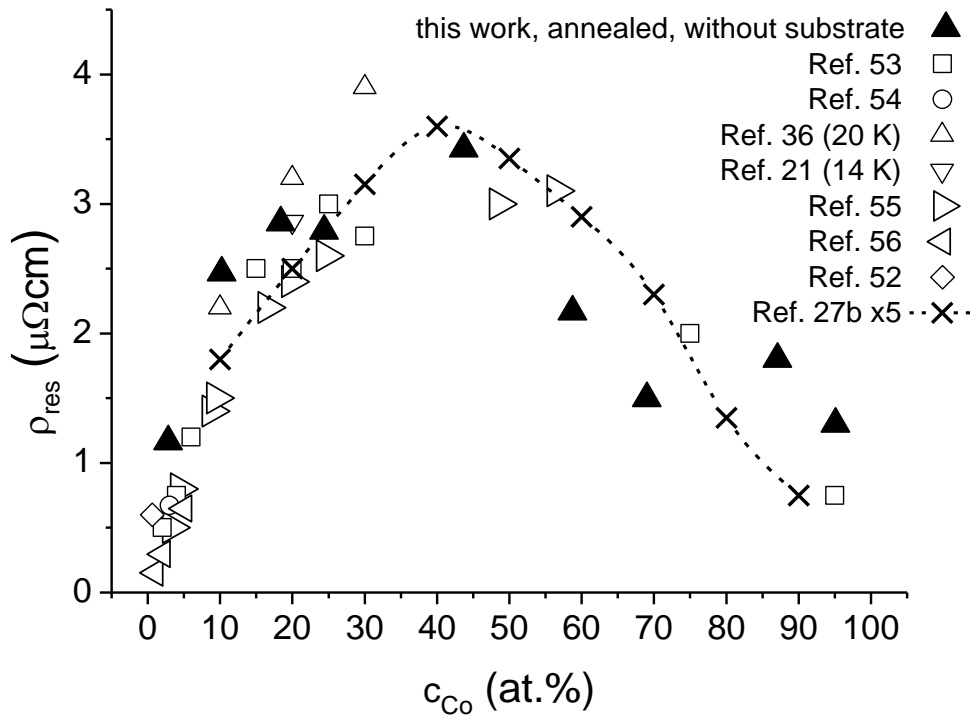

*Fig. 9* Residual resistivity data (▲) of Ni-Co alloys (annealed, without substrate) as a function of the Co-content. The open symbols refer to the literature data on metallurgically processed Ni-Co alloys as indicated in the legend. The results of an available theoretical calculation [27b] are shown by the crosses (×), by displaying 5 times the calculated values.

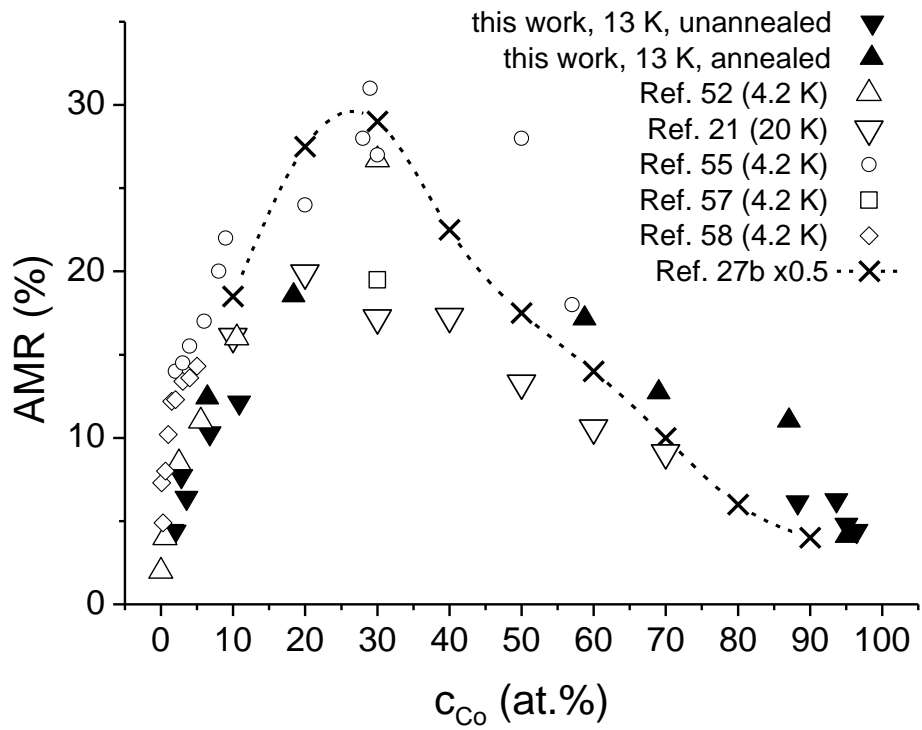

*Fig. 10* Low-temperature *AMR* data (▲) of annealed and unannealed ED Ni-Co alloys. The open symbols refer to the literature data on metallurgically processed Ni-Co alloys. The results of an available theoretical calculation [27b] are shown by the crosses (×), by displaying 0.5 times the calculated values.